# Multimodal, Multi-Disease Medical Imaging Foundation Model (MerMED-FM)

Yang Zhou PhD*[1]

Chrystie Wan Ning Quek MBBS*[2]

Jun Zhou PhD [1]

Yan Wang PhD [1]

Yang Bai PhD [1]

Yuhe Ke MD [3, 4]

Jie Yao MD [2]

Laura Gutierrez MD [2,5]

Zhen Ling Teo FRCOphth [2]

Darren Shu Jeng Ting FRCOphth PhD [2,6,7,8,9]

Brian T Soetikno MD PhD [10]

Christopher S. Nielsen PhD [11]

Tobias Elze PhD [12]

Zengxiang Li PhD [5]

Linh Le Dinh PhD [1]

Lionel Tim-Ee Cheng FRCR [4,13]

Tran Nguyen Tuan Anh FRCR [4,13]

Chee Leong Cheng FRCPath [14]

Tien Yin Wong MD PhD [2,15,16,17]

Nan Liu PhD [19,20,21]

Iain Beehuat Tan FAMS PhD [15,22,23]

Tony Kiat Hon Lim FRCPath[14]

Rick Siow Mong Goh PhD[1]

Yong Liu PhD[†1]

Daniel Shu Wei Ting MD PhD [†2,4,5,6]

Affiliations:

1. Institute of High Performance Computing (IHPC), A*STAR
2. Singapore National Eye Centre, Singapore Eye Research Institute, Singapore
3. Department of Anesthesiology, Singapore General Hospital.
4. Duke-NUS Medical School, Singapore, Singapore
5. Singapore Health Services, Artificial Intelligence Office
6. Ophthalmology and Visual Sciences Academic Clinical Program, Duke-NUS Medical School, Singapore

7. Academic Unit of Ophthalmology, Institute of Inflammation and Ageing, College of Medical and Health, University of Birmingham, Birmingham, UK.
8. Birmingham and Midland Eye Centre, Sandwell and West Birmingham NHS Trust, Birmingham, UK.
9. Academic Ophthalmology, School of Medicine, University of Nottingham, Nottingham, UK.
10. Byers Eye Institute, Stanford University School of Medicine, Palo Alto, California.
11. Cumming School of Medicine, University of Calgary, Calgary, Alberta, Canada
12. Massachusetts Eye and Ear, Department of Ophthalmology, Harvard Medical School, USA
13. Department of Diagnostic Radiology, Singapore General Hospital, Singapore
14. Department of Anatomical Pathology, Singapore General Hospital, Singapore
15. School of Clinical Medicine, Beijing Tsinghua Changgung Hospital, Tsinghua Medicine, Tsinghua University, Beijing, China
16. Beijing Visual Science and Translational Eye Research Institute, Beijing Tsinghua Changgung Hospital Eye Center, Tsinghua Medicine, Tsinghua University, Beijing, China
17. School of Biomedical Engineering, Tsinghua Medicine, Tsinghua University, Beijing, China
18. Division of Medical Oncology, National Cancer Centre Singapore, Singapore
19. Centre for Quantitative Medicine, Duke-NUS Medical School, Singapore, Singapore.
20. Programme in Health Services and Systems Research, Duke-NUS Medical School, Singapore
21. NUS AI Institute, National University of Singapore, Singapore, Singapore
22. Department of Radiology, University of Calgary, Calgary, Alberta, Canada
23. Data and Computational Science Core, National Cancer Centre Singapore, Singapore, Singapore.

*These authors contributed equally to this work as joint first authors
†Corresponding author

**Corresponding Author:**
Assoc Prof. Daniel Ting Shu Wei
Director, AI Office, Singapore Health Service
Singapore National Eye Centre, Singapore Eye Research Institute, Singapore
The Academia, 20 College Rd, Level 6 Discovery Tower, Singapore 169856
Email address: daniel.ting@duke-nus.edu.sg

**Financial Disclosure**: This study was supported by grants from the National Medical Research Council, Singapore (MOH-000655-00 and MOH-001014-00), Duke-NUS Medical School (Duke-NUS/RSF/2021/0018, 05/FY2020/EX/15-A58, and 05/FY2022/EX/66-A128), the Agency for Science, Technology and Research, Singapore (A20H4g2141 and H20C6a0032), Research to Prevent Blindness and NIH P30 EY003790.

**Availability of Data**: Additional data may reasonably be requested from the corresponding author

Main text word count (excluding abstract, methods, references, figure legends): 3952 words
Abstract word count: 150 words

**Abstract**

Current artificial intelligence models for medical imaging are predominantly single modality and single disease. Attempts to create multimodal and multi-disease models have resulted in inconsistent clinical accuracy. Furthermore, training these models typically requires large, labour-intensive, well-labelled datasets. We developed MerMED-FM, a state-of-the-art multimodal, multi-specialty foundation model trained using self-supervised learning and a memory module. MerMED-FM was trained on 3.3 million medical images from over ten specialties and seven modalities, including computed tomography (CT), chest X-rays (CXR), ultrasound (US), pathology patches, color fundus photography (CFP), optical coherence tomography (OCT) and dermatology images. MerMED-FM was evaluated across multiple diseases and compared against existing foundational models. Strong performance was achieved across all modalities, with AUROCs of 0.988 (OCT); 0.982 (pathology); 0.951 (US); 0.943 (CT); 0.931 (skin); 0.894 (CFP); 0.858 (CXR). MerMED-FM has the potential to be a highly adaptable, versatile, cross-specialty foundation model that enables robust medical imaging interpretation across diverse medical disciplines.

Medical imaging is increasingly fundamental to modern day medicine, playing a critical role in screening, triaging, diagnosis, prognosis and treatment guidance. Modalities such as radiological images (e.g., computed tomography [CT] scans), histopathological, dermascopy and ophthalmology images are essential in clinical workflows.[1] For example, histopathology remains gold standard in carder diagnosis, while ophthalmic imaging is critical for identification of sight-threatening disorders like age-related macular degeneration.[2,3]

While artificial intelligence (AI) and deep learning (DL) have revolutionized medical imaging tasks, most models remain narrow in scope, trained on a single imaging modality (e.g., colour fundus photography [CFP]) for specific tasks (e.g., diabetic retinopathy). This fragmented approach limits clinical utility as real-world diagnostics often require the physician to integrate information from multiple imaging sources. For instance, cancer diagnosis often necessitates both radiological imaging for staging and histopathological confirmation.[4] Despite this clinical reality, most leading foundational models (FM), such as ELIXR, MaCo, Rad-DINO and CXR Foundation, for chest X-ray (CXR) interpretation; Merlin and 3D Foundation AI model and Foundation model, for cancer imaging biomarkers for CT scans; USFM, for ultrasound (US) scans; UNI, CONCH and Virchow2, for pathology; RETFound, EyeFound and VisionFM, for ophthalmology imaging analysis; and PanDERM, for dermatology, are predominantly single modality and single-speciality. [5-19]

Developing a true multimodal, multispecialty and multi-disease FM is challenging yet essential. First, the difficulty lies in integrating diverse imaging data, as similar-looking features across modalities can confuse learning and dilute performance.[4,18] Second, training is constrained by the need for large, high-quality labeled datasets, which are often siloed and difficult to access. Third, many high-performing models, such as BiomedCLIP, use proprietary data, limiting real-world use.[20,21] Crucially, having an all-in-one multispecialty model offers practical advantages for clinical deployment, easing integration with hospital systems, reducing maintenance demands, and supporting consistent oversight of AI use in care workflows.

In this study, we propose MerMED-FM, a single, multimodal, multi-disease, multi-specialty FM to directly address these gaps. Firstly, MerMED-FM diagnoses various imaging modalities within a single architecture, enabling broad generalization and cross-specialty knowledge transfer. MerMED-FM is trained on 3.3 million medical images, spanning more than ten specialities (i.e. eye, lung, adrenal, spleen, kidney, bladder, prostate, liver, gallbladder, pancreas, colon, ovarian, uterine, bone, thyroid, skin, vessels, etc) and multiple modalities including chest X-ray (CXR), CT, ultrasound (US), histopathology slides, CFP, optical coherence tomography (OCT) and dermatology images. Secondly, MerMED-FM is trained using self-supervised learning (SSL), enabling efficient learning from large amounts of unlabelled data without expert annotation. Thirdly, it introduces a teacher-student iterative training system and a memory-based module to help the model learn more accurately from limited data and stay consistent across tasks, making it more reliable and useful in real-world healthcare settings.

Our key results show that MerMED-FM outperformed or matched learning specialty-specific models across multiple tasks. It showed strong performance in diagnosing lung cancer and COVID-19 pneumonia from CT scans, pneumonia and pneumothorax from CXR and various retinal diseases from OCT and CFP. It also performed comparably on US scans and histopathological cancer diagnosis and was found to be data efficient. By bridging traditionally siloed and isolated imaging domains, MerMED-FM aims to be an effective unsupervised, multimodal, multi-disease, multi-specialty FM with the potential to enhance diagnostic accuracy, streamline clinical workflows, and facilitate a patient-centered, cross-specialty approach in the complex hospital and healthcare environment.

**RESULTS**

We systematically evaluated MerMED-FM across multiple disease types and imaging modalities, comparing its to leading multispecialty and single-modality FMs. Performance was assessed across seven medical imaging types, covering CXR, CT, US, histopathology, CFP, OCT and dermoscopy photographs (Fig. 1e). Twenty-five publicly available datasets were used to assess its performance in diagnosing a range of conditions, including eye diseases (diabetic retinopathy, glaucoma, age-related macular degeneration, etc.), lung diseases (tuberculosis, COVID-19, pneumothorax, etc.), malignancies (colorectal, breast, lung cancer, etc.), and skin diseases (actinic keratosis, basal cell carcinoma, melanoma, etc.) (Supplementary Table 1). MerMED-FM was benchmarked against general-domain FM DINO, multispecialty FM BiomedCLIP and respective single-modality FMs (Supplementary Table 2). [7,8,10,17-19,22,23] All results are based on training with the full fine-tuning dataset (100%) unless otherwise stated.

MerMED-FM achieved the highest overall performance, with a mean AUROC of 0.935, outperforming BiomedCLIP (0.919) and Dino (0.933), and showed the best sensitivity (MerMED-FM: 78.3%, DINO: 77.2%, BiomedCLIP: 72.2%) and specificity (MerMED-FM: 91.8%, DINO: 91.3%, BiomedCLIP: 89.7%). Results are detailed in Table 1, with comparisons in Tables 2, 3 and Supplementary Table 3. Comparative radar plots are shown in Fig. 1a, normalized for clarity using the following formula:

$$Normalised\ value = 0.2 + \frac{(x - mix(x)) \times 0.8}{\max(x) - \min(x)}$$

**Performance by modality and disease type**

**Radiological image recognition and classification**

MerMED-FM was evaluated across ten CT, CXR and US datasets.[24-32] It was found to be the leading model in detecting lung carcinomas and COVID-19 pneumonia on CT scan slices, recording an overall mean AUROC of 0.975. It had the highest sensitivity of 90.3% (mean sensitivities: BiomedCLIP = 81.1%, Merlin = 81.6%, Dino = 97.7%) and highest specificity of 98.2% (mean specificities: BiomedCLIP = 95.5%, Merlin = 96.0%, Dino = 97.7%).[7,22,23] For identifying lung carcinomas such as adenocarcinoma and squamous cell carcinoma on chest CT scans, the model achieved a mean AUROC of 0.981, and the highest mean sensitivity of 92.4% and specificity of 98.2% compared to BiomedCLIP, Merlin and Dino.[7,22,23] In particular, results showed that MerMED-FM (AUROC = 0.974, CI: 0.964-0.984) surpassed the leading CT scan domain-specific model, Merlin (AUROC = 0.940, CI: 0.926-0.953), by a margin of 3.61% in diagnosing lung cancer on the IQ-OTHNCCD dataset.[7,26] This difference was not only found to be statistically significant (p<0.01, T = 6.02), but also clinically relevant with a Cohen's d of 3.51. Similarly, it outperformed BiomedCLIP by 0.93% (p<0.01, T = 0.94, Cohen's d = 0.951) and Dino by 1.09% (p=0.036, T = 3.10, Cohen's d = 1.49) in this domain.[22,23] This was consistent on the chest-ctscan-imaging dataset, where it performed 0.652% better than Merlin (p value = 0.017, T statistic 3.92, Cohen's d = 2.52. While Dino (AUROC = 0.990, CI: 0.986-0.994) had a marginally better performance by 0.23% than MerMED-FM (AUROC = 0.988, CI: 0.985-0.990) on the chest-ctscan-imaging dataset for recognizing lung cancer on CT scans, this difference was not statistically proven (p=0.137, T = 1.86).[18,23,25]

Similarly, superior results were found for MerMED-FM in the recognition of COVID-19 pneumonia on CT scans, with a mean AUROC of 0.999, sensitivity of 98.6% and specificity of 99.0%. Compared to Merlin (AUROC = 0.987, CI: 0.987-0.993), MerMED-FM (AUROC = 0.999, CI: 0.999-0.999) demonstrated a statistically significant improvement of 1.17% with a substantial effect of Cohen's d of 15.3 on the SARS-COV-2 dataset (p<0.01, T = 27.4).[7] Compared to the next best performing model, Dino (AUROC = 0.997, CI: 0.995-0.999), it was better by 0.237% (p=0.0372, T = 3.07).[23] For identifying COVID-19 pneumonia on CT scans on the iCTCF dataset, MerMED-FM outperformed Merlin by a statistically significant difference of 0.0781% (p value < 0.01, T statistic = 39.0, Cohen's d = 6.37).[7,29]

When tested on interpreting CXRs, MerMED-FM had a mean AUROC of 0.908, sensitivity of 78.9% (mean sensitivities: BiomedCLIP = 78.2%, RadDINO = 79.9%, Dino = 79.1%) and specificity of 86.9% (mean specificities: BiomedCLIP = 83.8%, RadDINO = 87.1%, Dino = 85.5%).[10,22,23] MerMED-FM (AUROC = 0.890, CI: 0.889-0.891) was found to be better than the top-performing CXR-specialised model, RadDINO (AUROC = 0.793, CI: 0.785-0.802), in recognizing pneumonia by a notable difference of 12.2%.[10] This translates to a Cohen's d of 20.0 which was shown to be strongly statistically significant (p<0.01, T = 29.8). For the diagnosis of pneumothorax on CXRs, MerMED-FM (AUROC = 0.934, CI: 0.932-0.937) outperformed RadDINO (AUROC = 0.903, CI: 0.893-0.913) by a statistically significant difference of 3.45% and a large effect of Cohen's d 5.20 (p<0.01, T statistic 7.92).[10] All four models were on par for identifying tuberculosis on CXRs (MerMED-FM's AUROC = 0.999, CI: 0.999-1.000; RadDINO's AUROC = 0.999, CI: 0.999-0.999, BiomedCLIP's AUROC = 1.00, CI: 1.00-1.00; Dino's AUROC = 0.999, CI: 0.999-0.999; all p>0.05).[10,21-23,30] MerMED-FM was found to narrowly tail the best performing model in CXR interpretation, Dino, in diagnosing pneumothorax by 0.399% (BiomedCLIP's AUROC = 0.928, CI: 0.925-0.931; p<0.01, T = -4.98, Cohen's d = -2.45).[22,23]

MerMED-FM was shown to be superior or comparable to the leading single-specialty model, USFM, in the identification of breast cancer on US scans across three datasets.[31,32] It achieved a mean AUROC of 0.951 (mean AUROCs: BiomedCLIP = 0.909, USFM = 0.939, Dino = 0.956), the highest specificity of 98.6% (mean specificities: BiomedCLIP = 87.0%, USFM = 88.7%, Dino = 90.1%).[17,22,23] Meanwhile, USFM demonstrated a mean AUROC of 0.939, with a percentage difference of 1.28% as well as a sensitivity of 86.2% and specificity of 88.7%.[17] On the BUSI dataset, MerMED-FM scored an AUROC of 0.973 (CI: 0.967-0.979), higher than USFM's 0.966 (CI: 0.963-0.969).[17,32] This had a moderately large effect with Cohen's d score of 2.00 which was statistically significant (p=0.0328, T = 3.20). MerMED-FM was found to be comparable to BiomedCLIP and Dino as their differences were shown to be statistically insignificant (p>0.050).[22,23]

**Histopathology slide interpretation**

Evaluation of MerMED-FM's performance on histopathology patches was conducted on two datasets.[33-35] It has excellent performance for detecting breast cancer and various other types of pathology. A mean AUROC of 0.999 (means AUROCs: BiomedCLIP = 0.988, UNI = 0.999, DINO = 0.999), the highest mean sensitivity of 99.5% (mean sensitivities: BiomedCLIP = 93.1%, UNI = 99.0%, DINO = 98.7%) and the highest mean specificity of 99.8% (mean specificities: BiomedCLIP = 97.0%, UNI = 99.2%, DINO = 99.4%) were achieved by MerMED-FM. MerMED-FM scarcely outperformed UNI by 0.120% in diagnosing breast carcinoma while performing worse than UNI by 0.0240% in diagnosing various tissue types on histopathology slides.[19,22,23] However, differences were not found to be statistically significant (p>0.050), highlighting the equivalent capabilities of the model with this leading histology-specific model.[19] While UNI was trained on over 100 million pathology patches, MerMED-FM's self-supervised training with fewer images and no label still showcased remarkable efficacy.[19]

**Ocular disease diagnosis**

We evaluated the performance of MerMED-FM on seven ophthalmological imaging datasets in diagnosing eye conditions using OCT and CFP images.[36-43] MerMED-FM was found to be the best performing model in identifying various retinal diseases, such as diabetic retinopathy (DR), age-related macular degeneration (AMD), epiretinal membrane (ERM), and retinal vein and arterial occlusions on OCT. It attained an AUROC of 0.998 (confidence interval [CI]: 0.997-0.999) on the OCTID dataset, significantly outperforming RETFound, the leading ophthalmology model, which recorded an AUROC of 0.960 (CI: 0.954-0.966) by a statistically significant difference of 3.92% ($p<0.001$, T = 17.9) and large effect with a Cohen's d of 10.6.[8,38] On the OCTDL dataset, it also performed better than RETFound by 0.491% and Cohen's d of 3.01 ($p < 0.01$, T = 5.60) where achieved an AUROC of 0.911 (CI: 0.990-0.992) compared to RETFound's 0.986 (CI: 0.984-0.989). On the other hand, MerMED-FM performed slightly worse than BiomedCLIP on the OCTDL dataset by 0.292% ($p < 0.01$, T = -9.26, Cohen's d = -3.68).[22] Overall, MerMED-FM demonstrated the highest mean AUROC of 0.988, compared to RETFound's 0.974, BiomedCLIP's 0.981 and Dino's 0.985. While it had the highest mean sensitivity of 92.8% (mean sensitivities: BiomedCLIP = 89.5%, RETFound = 80.7%, Dino = 91.1%), it had the lowest specificity of 85.9% (mean specificities: BiomedCLIP = 95.5%, RETFound = 95.1%, Dino = 95.9%) on OCT imaging tasks.[8,22,23]

For diagnosing various ocular conditions on CFP, MerMED-FM was found to have an overall mean AUROC of 0.894, and the highest mean specificity of 88.8% (mean specificities: BiomedCLIP = 86.4%, RETFound = 87.5%, Dino = 88.3%) and highest mean sensitivity of 64.3% (mean sensitivities: BiomedCLIP = 54.1%, RETFound = 59.4%, Dino = 62.1%).[8,22,23] MerMED-FM showed unrivalled results in detecting glaucoma on CFP. It achieved an AUROC of 0.957 (CI: 0.953-0.961), significantly superseding RETFound which recorded an AUROC of 0.937 (CI: 0.935-0.939) on the Glaucoma Fundus dataset.[8,39,42,44] This translates to a 2.14% improvement with a statistically (p<0.01, T = 12.6) and practically significant difference (Cohen's d = 7.40). MerMED-FM was also better than one of the leading multispecialty models, BiomedCLIP (AUROC = 0.919, CI: 0.916-0.921), at recognizing glaucoma on CFP with a statistically significant improvement of 4.14% (p<0.01, T = 17.7) and Cohen's d of 13.5 on the Glaucoma Fundus dataset and a substantial 9.72% (p<0.01, T=7.84) and Cohen's d of 5.39 on the PAPILA dataset.[21,22] While MerMED-FM (AUROC = 0.836, CI: 0.820-0.852) had a slightly inferior performance of 1.25% compared to RETFound (AUROC = 0.847, CI: 0.834-0.860) on the PAPILA dataset, this difference was not statistically significant (p value=0.0573).[8,40]

For diagnosing DR and other retinal diseases, MerMED-FM achieved a mean AUROC of 0.852 and 0.972, and high specificity of 90.8% and 99.1% respectively. On the JSEIC dataset, MerMED-FM (AUROC = 0.996, CI: 0.993-0.998) was found to be better than RETFound (AUROC = 0.990, CI: 0.989-0.990) by 0.63% and Cohen's d of 4.36 which was statistically significant (p=0.002, T = 7.24), superior to BiomedCLIP (AUROC = 0.989, CI: 0.987-0.991) by 0.740% and Cohen's d of 4.14 and on par with Dino (AUROC = 0.994, CI: 0.991-0.998) where their difference was statistically insignificant (p=0.183).[8,23,41] Conversely, Dino (AUROC = 0.987, CI: 0.977-0.997) was better than MerMED-FM (AUROC = 0.949, CI: 0.941-0.957) by 4.02% on the CRFO-v4 dataset (p<0.001, T = 11.3, Cohen's d = 5.39).[23,36] For DR on CFP, MerMED-FM (AUROC = 0.768, CI: 0.753-0.783) mostly had poorer results than the other datasets.

**Dermatological lesion identification**

Four datasets were utilized for assessing MerMED-FM's performance in diagnosing dermatological diseases from clinical and dermoscopic images.[45-47] Overall, MerMED-FM achieved a mean AUROC of 0.931, sensitivity of 75.9% (mean sensitivities: BiomedCLIP = 71.7%, PanDERM = 79.0%, Dino = 75.1%) and specificity of 92.1% (mean specificities: BiomedCLIP = 90.2%, PanDERM = 92.4%, Dino = 91.3%). MerMED-FM (AUROC = 0.962, CI: 0.961-0.963) was found to closely follow PanDERM's performance by 0.81% on the Dermnet dataset at a T statistic of 14.34 which was statistically significant (p<0.01) (AUROC = 0.970, CI: 0.968-0.972).[18] Additionally, on the PAD-UFES-20 dataset, PanDERM (AUROC = 0.947, CI: 0.943-0.951) was 1.80% better than MerMED-FM (AUROC = 0.930, CI: 0.918-0.942) with a Cohen's d of -2.44 (p=0.020, T = -3.78).[18,46] PanDERM's training dataset varies in size and diversity, leveraging millions of domain-specific samples, while MerMED-FM demonstrated comparable performance using balanced, multi-domain training with fewer but more diverse datasets.[18] These models were fine-tuned on datasets within their respective domains.

**Clinical applicability of MerMED-FM**

To determine MerMED-FM's clinical deployability and effectiveness on diverse real-life diagnostic tasks, it was evaluated on six ophthalmological, two radiographic and one histopathological datasets from a local tertiary hospital cluster in Singapore (Supplementary Table 4). These findings are presented in detail in Tables 2, 3 and Supplementary Table 3.

MerMED-FM (AUROC = 0.808, CI: 0.802-0.814) outperformed Dino (AUROC = 0.788, CI: 0.783-0.793) and BiomedCLIP (AUROC = 0.760, CI: 0.752-0.768) and was comparable to RadDINO (AUROC = 0.807, CI: 0.796-0.818) in diagnosing various respiratory diseases on the local AIMx-CXR dataset.[10,22,23] It was on par with comparator models in diagnosing various hepatic conditions such as hepatocellular carcinoma, other liver malignancies, hemangioma, abscess, cyst, focal nodular hyperplasia and benign lesions on the RAPIER Gastric dataset. MerMED-FM achieved an overall AUROC of 0.916 (CI: 0.901-0.931), compared to Merlin's 0.939 (CI: 0.932-0.946), BiomedCLIP's 0.953 (CI: 0.948-0.958) and Dino's 0.937 (CI: 0.915-0.960) with statistically insignificant differences ($p > 0.05$). [7,22,23]

Excellent performance was shown in its ability to identify diabetic macular edema on OCT with a AUROC of 0.975, sensitivity of 90.6% and specificity of 90.6%. For diagnosing diabetic macular edema on OCT on the DRCR (OCT) dataset, MerMED-FM (AUROC = 0.975, CI: 0.971-0.979) was comparable to RETFound (AUROC = 0.908, CI: 0.908-0.976) with a statistically insignificant difference of 0.15% ($p=0.385$, $T = -0.976$).[8]

For diagnosing diabetic macular edema on CFP on the DCDR dataset, MerMED-FM was the best performing model with an AUROC of 0.820 (CI: 0.810-0.830), compared to RETFound's 0.777 by 5.46% (CI: 0.770-0.785), BiomedCLIP's 0.722 by 6.20% (CI: 0.764-0.780) and Dino's 0.794 by 3.19% (CI: 0.792-0.797).[8,22,23] These differences were statistically significant ($p <0.01$, $T > 8$) with large effect sizes of Cohen's d 6.04, 6.31 and 6.61 respectively.[8,22,23] MerMED-FM (AUROC = 0.937, CI: 0.932-0.942) also outperformed RETFound (AUROC = 0.849, CI: 0.835-0.863) in identifying AMD on CFP by a statistically significant difference of 10.7% ($p<0.01$, $T = 14.1$) and great effect of Cohen's d 10.6.[8] On the other hand, MerMED-FM (AUROC = 0.816, CI: 0.808-0.923) lagged RETFound (AUROC = 0.837, CI: 0.832-0.842) by 2.59% in diagnosing glaucoma on CFP.[8] MerMED-FM demonstrated comparable performance to RETFound for recognizing DR and MMD on CFP.[8]

**Data efficiency and low-shot adaptability of MerMED-FM**

MerMED-FM's data efficiency was evaluated by comparing performance using 10%, 30%, and 50% of the available fine-tuning data (Fig. 2; Supplementary Table 5). MerMED-FM demonstrated strong adaptability limited supervision, maintaining high diagnostic accuracy even with a fraction of the data. When fine-tuned with only 50% of the data, it achieved an AUROC of 0.926 and specificity of 91.1%, close to the full-data benchmark (AUROC 0.948, specificity 93.5%).

Across several imaging modalities, MerMED-FM consistently outperformed or matched state-of-the-art single-modality and generalist FMs in low-data settings. On CT scans, it surpassed BioMed-CLIP and DINO in both AUROC and F1 scores at all data levels (10% to 100%) (Fig. 2(b)). In retinal imaging tasks (CFP and OCT), MerMED-FM outperformed RETFound, BioMed-CLIP, and DINO in AUROC and F1 scores, even with just 10% of fine-tuning data (Fig. 2(e)(f)).[8,22,23]

For US, MerMED-FM achieved the highest F1 score (0.708) using only 10% of training data—outperforming USFM (0.662), BiomedCLIP (0.665), and DINO (0.656) (Fig. 2(c)).[17,22,23] On CXR tasks, MerMED-FM matched or exceeded other models in AUROC across all data fractions, although it trailed Rad-DINO in F1 performance (Fig. 2(a)).[10]

For histopathology and dermatology tasks, MerMED-FM demonstrated competitive data efficiency compared to generalist models (BiomedCLIP and DINO), but still fell short of specialist models like UNI (histopathology) and PanDERM (dermatology) (Fig. 2(d)(g)).[10,18,19,23]

**DISCUSSION**

The current state-of-the-art FMs in medical imaging are largely based on siloed and fragmented approaches, not reflecting real-world clinical practice in complex multi-disciplinary hospital environments. Most FMs focus on a single imaging modality (e.g. CFP), for specific diseases within a certain specialty (e.g., diabetic retinopathy in ophthalmology). In reality, conditions, such as diabetes mellitus and cancer, often require integration of diverse imaging modalities to evaluate systemic complications. We present MerMED-FM, a data-efficient multimodal, multi-disease, multi-specialty, vision-only FM trained with SSL, with the potential to enhance screening, diagnostic accuracy, streamline clinical workflows, and facilitate a patient-centered, cross-specialty approach in the complex hospital and healthcare environment.

Unlike leading models such as RETFound, Rad-DINO, Merlin, USFM, PanDERM, and UNI, which excel in their respective domains but require separate pipelines, MerMED-FM unifies the interpretation of diverse imaging types and possesses the multi-specialty adaptability needed to support comprehensive diagnostic reasoning.[7,8,10,17-19] This reduces the need for separate models, cuts computational overhead and enhances clinical efficiency. For example, it can assess breast US scans and pathology slides for breast cancer in parallel or separately evaluate colorectal cancer pathology slides alongside CT scans for metastatic staging. By handling different types of medical images within one model without requiring separate models or heavy reliance on language-based inputs, MerMED-FM represents a major step toward integrating AI meaningfully into real-world clinical workflows.

The key challenge of developing a multimodal multi-speciality model stemmed from harmonising heterogenous data, varying in resolutions, scales, and structures, without compromising performance on specific tasks.[18] MerMED-FM presents a technological breakthrough and circumvents through three main strategies: (i) modality-specific preprocessing and normalisation to embed inputs in a shared latent space, (ii) a memory-based training framework is spearheaded to enable retention and cross-modal understanding of clinically relevant features, and (iii) balanced modality- and specialty-aware sampling to avoid domain dominance during training to ensure equity and generalisability to underrepresented and rare diseases. Together, these features make MerMED-FM a robust, versatile tool that supports accurate diagnosis and decision-making across the complex, multimodal landscape of modern medicine.

Data scarcity and annotation burden remain major bottlenecks in medical AI. Central to MerMED-FM's development is the use of SSL, differentiating itself from existing models, learning meaningful representations from unlabelled data and significantly reducing reliance on labeled data.[21,22,48,49] Its memory module supports progressive learning from new data, similar to how clinicians develop pattern recognition skills over years of experience. This approach transforms previously untapped medical imaging data into a valuable resource for AI-driven diagnosis and decision-making.

MerMED-FM demonstrated excellent and robust capabilities across multiple imaging modalities. It outperforms or closely matches specialty-specific models in key domains. First, it outperformed all comparator models in diagnosing lung cancer and COVID-19 pneumonia on chest CT scans, including Merlin, a vision language FM dedicated for CT scan interpretation. Second, it was found to be better than RadDINO, a self-distilled vision transformer for CXRs, in detecting pneumonia and pneumothorax on CXRs.[10] Third, it had comparable performance to USFM and UNI in diagnosing breast cancer and malignancies on pathology slides (breast, colorectal, and lung cancer) respectively.[17,19] Fourth, it outperformed comparator models in identifying various retinal diseases on OCT and glaucoma on CFP, MerMED-FM surpasses RETFound, an SSL-based model for ocular disease recognition.[8] Last, its performance under low-data regimes highlights its data-efficiency and potential for rapid deployment across new clinical tasks and imaging modalities with minimal annotation effort.

Compared to emerging multimodal models such as HealthGPT, which rely on language inputs and written task instructions, MerMED-FM offers a vision-only alternative that is more useful in clinical scenarios where medical images come unannotated or processed.[50] It learns directly from image patterns themselves, without needing specific instructions. Unlike multispecialty models, such as BiomedCLIP, UniMed-CLIP, BIOMEDICA and MedImageInsight, that rely on image-text pairs, MerMED-FM is purely image-based.[18,21,22,48] This makes the comparison uneven and may artifactually offer those multispecialty models an advantage in statistical analysis. This distinction is crucial because in clinical practice, images are acquired before textual descriptions are generated. Models requiring text inputs already assume physician involvement, limiting their utility in triaging and real-time decision-making. MerMED-FM aligns more closely with real-world workflows, making it a more practical solution for hospital settings.

Potential clinical applications of MerMED-FM are multi-fold where its versatility was demonstrated through its evaluation on a local dataset from a tertiary health cluster (Supplementary Table 4). Firstly, it may be deployed for triage by providing real-time preliminary reporting of medical images, especially in emergency and critical care settings. Secondly, it may provide diagnostic support by providing insight to clinically ambiguous findings. Thirdly, it may be used for multitask diagnosis, such as in a patient with both diabetic retinopathy and kidney disease. Separate diagnostics tools would not be needed to interpret the CFP and US kidneys. These capabilities offer opportunities to reduce diagnostic delays, streamline imaging interpretation and support treating for junior clinicians. A single unified model could reduce the logistical and maintenance burden of managing multiple domain-specific tools across departments.

Despite its strengths, several limitations remain. MerMED-FM did not outperform single-modality FMs in pathology and dermatology tasks. It currently does not support volumetric imaging, instead relying on 2D slices. MerMED-FM has yet to be evaluated for synergistic multimodal reasoning within individual patients. For instance, evaluating both both chest CT scans and histopathology to arrive at a unified diagnosis of lung adenocarcinoma. Such integration may enhance diagnostic precision and support complex decision-making.

To further enhance its clinical impact, several future directions are proposed. Beyond classification, MerMED-FM can evolve to support prognostication (i.e., including longitudinal data for regression using long short-term memory frameworks), segmentation (i.e., delineating tumor margins), and report generation. The integration of individual patient's clinical history or examination findings may improve the diagnostic accuracy. The inclusion of synthetic data from generative adversarial networks or diffusion models could augment training and improve robustness. Regulatory and ethical concerns must be addressed to facilitate MerMED-FM's real-world adoption. MerMED-FM represents a stepping stone to unifying imaging across specialties and has the potential to enhance diagnostic accuracy, improve workflow efficiency, and eventually integrate AI into everyday clinical practice.

**METHODS**

The training pipeline of MerMED-FM follows the process outlined in Fig. 3a. Firstly, multiple augmented views are generated from an input image. Second, each view is encoded separately using both student and teacher encoders, producing corresponding vector representations. Third, these representations are compared with stored imaging representations in memory to compute similarity distributions. Finally, the similarity distributions for different views are checked to ensure that they align with those stored in memory, reinforcing consistency.

**Pre-Training Dataset**

To develop MerMED-FM, we utilised publicly available unlabelled datasets for large-scale pretraining across diverse medical imaging modalities and specialties. Approximately 3.3 million images were used, spanning seven modalities and over ten specialties. The dataset includes 292,353 CT slices, 713,931 CXR, 389,885 ultrasound frames, 333,700 CFP, 176,719 OCT slices, 1,017,712 pathology patches, and 401,059 dermoscopic images (Fig. 1b, c). These images cover radiology, oncology, urology, gynecology, hepatology, nephrology, pulmonology, ophthalmology, dermatology, and pathology (Supplementary Table 6). For volumetric modalities like CT and US, we used provided datasets that were originally provided as two-dimensional slices without additional preprocessing, applied consistently across training and evaluation. The study was approved by the Singapore Eye Research Institute Ethics Committee and the Singapore Health Services Centralized IRB, in accordance with the Declaration of Helsinki.

**Overcoming Challenges in Building a Multispecialty, Multimodal FM**

Building a FM that can interpret multiple imaging modalities across diverse clinical specialties introduces several fundamental challenges. Firstly, medical images differ significantly in visual structure, noise characteristics and clinical semantics, making it difficult to unify them under a single encoder without sacrificing modality-specific fidelity. Secondly, representations learned from one modality may interfere with those from another when trained jointly, degrading stability and feature learning, especially in early training stages. Thirdly, the risk of catastrophic forgetting exists when it loses its ability to retain knowledge from previously seen modalities, particularly in continual or mixed-domain training setups. MerMED-FM tackles these challenges through a combination of SSL and memory-based regularization.

**MerMED-FM Employs SSL and a Memory Module**

The memory-augmented SSL approach in MerMED-FM enables effective joint learning from heterogeneous medical imaging data. SSL allows the exploitation of large amounts of unlabeled data to learn general representations by solving pretext tasks. It harnesses the inherent pattern and structure within unlabelled data, mirroring the human learning process by extracting meaningful representations without the need for labelled data.[51,52] Unlike existing single-modality models, MerMED-FM integrates multiple imaging modalities through joint modelling methods (Fig. 3b). Central to this approach is a dual-network teacher-student framework where a slowly updated teacher model guides a student encoder during training. This architecture stabilises training by enforcing semantic consistency across different views, helping the student learn robust and transferable representations without abrupt updates.

A key strategy in MerMED-FM is its dynamic memory module which stores compact representations from past training samples across modalities. This was inspired by how humans use memory to contextualise new observations. MerMED-FM compares current image views to previously encountered concepts stored in memory. For instance, when comparing two CXRs from the same patient at different time points, the memory module identifies shared features while recognising new abnormalities. New image views are not only aligned with each other, but also with relevant representations stored in memory. This enforces temporal and inter-view consistency, maintains alignment across specialties and provides historical context to prevent forgetting.

The memory module serves two primary purposes. Firstly, this stabilizes the SSL as the memory acts as a reference, preventing mode collapse and ensuring consistent representations across diverse medical imaging modalities. Secondly, memory improves representations as it encourages consistency between unlabelled images and possibly labelled concepts stored in memory by comparing their features through non-parametric methods. Unlike traditional contrastive learning methods that rely on memory queues to mine negatives or positives, MerMED-FM aligns current inputs (views of an image) with previously stored concepts (representations from earlier iterations) to ensure consistency and meaningful connections between unlabelled and labeled data.

The memory updates iteratively, with new representations replacing older ones using a First-In, First-Out (FIFO) mechanism to ensure that the memory remains relevant. The memory module also regularizes learning using a stochastic partition strategy. By forcing representations to generalize rather than overfit specific patterns, MerMED-FM ensures robust and unbiased learning across modalities. The memory operates dynamically, updating stored representations iteratively, thus balancing historical context with the latest observations. At a high level, memory allows for three crucial processes: (1) acquisition of new information (encoding), (2) information retention over time (storage), and (3) retrieval. Through these processes, we can make sense of our present and take informed actions based on past observations.

The model also incorporates a balanced modality-specialty method of sampling, which ensures equal representation across imaging types and clinical domains within each batch. This prevents data imbalance from skewing learning and avoids overfitting to dominant modalities like CXR, improving generalisation to underrepresented domains. To prevent overfitting to recent data, MerMED-FM also employs a random retrieval mechanism when accessing memory. This strategy balances the influence of historical and recent representations, ensuring robust and unbiased learning. For example, when analyzing two views of a CXR, MerMED-FM assigns high similarity scores to related representations (e.g., similar lung patterns) and low scores to unrelated ones, such as features from other modalities. This ensures consistent learning across all views. (Fig. 3a)

**Model Architecture**

The joint-embedding teacher-student architecture is based on Vision Transformers (ViTs) as the backbone. Twelve augmented views were generated per training iteration, including two global views (224 × 224 pixels each) and ten local views (96 × 96 pixels each), using techniques like color jittering, Gaussian blur, solarization, and random cropping, following protocols established by Grill et al.[53]

Separate ViT encoders and projection heads were used for the student and teacher branches. Each projection head is a three-layer multilayer perceptron (MLP) with a hidden size of 2048 dimensions and Gaussian Error Linear Unit (GELU) activations, as proposed by Caron et al. (2020). [23,53] Only the student branch was directly updated using gradient descent, while the teacher branch was updated through an exponential moving average of the student's network weights. Each image view is encoded using the [CLS] token from the Transformer encoder. For instance, the ViT-B encoder mapped image views to 768-dimensional representation vectors, which were further projected to a lower 256-dimensional space and normalized to lie on a unit hypersphere.

The memory module (M) is a non-differentiable container that stores representations generated during training, operating using a FIFO update strategy. The memory size was fixed at K = 65536, as determined through ablation studies. During optimization, the view-memory similarity distribution was partitioned into memory blocks, each containing $N_b$ = 16384 representations.

Training was conducted using the AdamW optimizer[43] with a learning rate starting at $1 \times 10^{-5}$ and a global batch size of 1024. The learning rate decays according to a cosine schedule, converging to $1 \times 10^{-6}$ without a warm-up phase. Weight decay follows a cosine schedule from 0.04 to 0.4, as suggested by Caron et al.[16] The student temperature ($\tau_s$) was fixed at 0.1, while the teacher temperature ($\tau_t$) gradually increased from 0.04 to 0.07 over the first 30 epochs.

**Adaptation to Downstream Tasks**

To adapt MerMED-FM to downstream tasks, only the student encoder (ViT-base) was required. It extracts high-level visual features from medical images, which are processed by an MLP for task-specific predictions. The MLP takes these features as input and outputs the probability of disease categories. The category with the highest probability is selected as the final classification. The number of MLP output neurons corresponded to the number of disease categories in the task. The model (encoder and MLP) is then finetuned for individual downstream tasks using the respective training datasets.

To prevent overfitting, label smoothing was applied during training, softening ground-truth labels and regulating the output distribution. The training objective was to match the predicted categorical outputs with the ground-truth labels. The model was trained with a batch size of 16 for 50 epochs. The first 10 epochs used a learning rate warm-up from 0 to $5 \times 10^{-4}$, followed by a cosine annealing schedule reducing the learning rate to $1 \times 10^{-6}$ over the remaining 40 epochs. After each epoch, the model was evaluated on a validation set, and the checkpoint with the highest AUROC on the validation set was saved for further evaluation. This ensures optimal performance for both internal and external validation datasets.

The model was finetuned for individual downstream tasks using the respective training datasets as seen in Supplementary Table 1. This was similarly done for Biomed-CLIP, Dino and the single-modality comparator models. Finetuning using only 10%, 30% and 50% of the data was separately performed to assess the data efficiency of the model.

**Computational resources**

We used eight NVIDIA H100 (80 GB) GPUs, which takes around three days. For fine-tuning FMs to downstream tasks, we used an NVIDIA L40 (48 GB), which takes about 1-2 min per 1,000 images.

**Benchmarking Across Seven Medical Imaging Modalities**

We systematically evaluated MerMED-FM across seven medical imaging tasks, covering CT, CXR, US, CFP, OCT, histopathology whole slide images and dermoscopic photographs. Firstly, MerMED-FM was compared against general-domain FM, Dino, to highlight the benefits of domain-specific training.[23] Secondly, comparisons to top-performing single-modality FMs, RETFound, Rad-DINO, Merlin, USFM, UNI and PanDerm.[7,8,10,17-19] Thirdly, MerMED-FM was benchmarked against peer SOTA multispecialty FM, BiomedCLIP (Supplemantary Table 2). Model performance was primarily measured using AUROC for well-rounded representation.

A total of 25 widely recognized public datasets were used for evaluation as represented in Supplementary Figure 1. These datasets feature a myriad of diseases that are widely recognized for training or benchmarking domain-specific disease diagnosis were used for evaluation. Evaluation was conducted on annotated datasets with diabetic retinopathy (APTOS2019, IDRiD and MESSIDOR2 datasets), glaucoma (Glaucoma-Fundus and PAPILA datasets) and other retinal diseases (CRFO-v4 and JSIEC datasets) for CFP and OCT

(OCTDL and OCTID datasets). For CXR, datasets of annotated radiographs of pneumonia (RSNA dataset), tuberculosis (TBX11K dataset) and pneumothorax (SIIM dataset) were evaluated. For CT scans, annotated slices of lung carcinoma (Chest-CTscan-images and IQ-OTH/NCCD datasets), COVID-19 (SARS-COV-2 and iCTCF datasets) were utilized. US scans showing various breast cancer (BUSC, BUSI and BrEaST datasets) were used for comparisons. For histopathology, the models were evaluated on classification and segmentation tasks for WSI of colorectal adenocarcinoma (CRC-VAL-HE-7K dataset), breast cancer (BreakHis dataset) and other conditions (PanNuke dataset). Performance of the FMs on clinical skin and dermoscopic images were assessed using Derm7pt, PAD-UFES-20, Dermnet and HAM10000 datasets.

Seven real-world local datasets were used from a tertiary hospital cluster in Singapore to evaluate the clinical robustness and diverse capabilities of the model (Supplementary Table 4). MerMED-FM was then tested on seven datasets across CFP (DCDR [CFP] and FM [DR; AMD; Glaucoma; MMD]), OCT (DCDR [OCT]) and CXR (AIMx-CXR) sourced from a tertiary health cluster in Singapore (Supplementary Table 4). These datasets provide valuable perspective beyond the carefully curated public ones. While public datasets are often curated and standardized, local datasets reflect real-world variability in disease presentation, imaging equipment, demographics, and reporting standards. This allowed us to evaluate MerMED-FM's clinical robustness and generalizability in real-world settings.

To assess the statistical significance and consistency of MerMED-FM to its comparator models, five random independent runs were conducted across the various public and local test datasets. Paired t-tests were performed between MerMED-FM and the comparator models to obtain the p-value and T-statistic. The mean difference and percentage improvement and effect sizes (Cohen's d) were calculated (Table 3).

**Efficient learning from less data**

MerMED-FM demonstrates strong data efficiency, especially in the pretraining stage (Fig. 2). This efficiency stems from the model's ability to harmonize multiple modalities through a shared image encoder and stable semantic-level representation learning with memory module. This joint training strategy significantly reduces total data requirements while enabling effective cross-specialty generalization. By aligning new observations with previously stored concepts, the model reinforces semantic consistency and stabilizes training. This memory-guided learning allows the model to extract meaningful features even from limited data per modality, eliminating the need for redundant training pipelines.

## CONCLUSION

MerMED-FM represents a major step forward in applying AI across multiple medical imaging modalities. It outperforms existing single-modality and multi-specialty models in identifying ocular, lung, and systemic diseases. By leveraging SSL technique and memory module, the model is capable of processing multidimensional data with minimal dependence on labeled datasets. These findings underscore the feasibility of a unified AI model multispecialty diagnosis and lays a groundwork for future research in AI-driven diagnostic support.

**Acknowledgements**:

We would like to thank Prof Nigam Shah and Dr Curtis P Langlotz for their contribution and guidance on this the manuscript.

Writing of the manuscript was assisted by ChatGPT4o to correct grammatical errors and flow of the introduction and discussion. Results and ideas are original.

# Figures

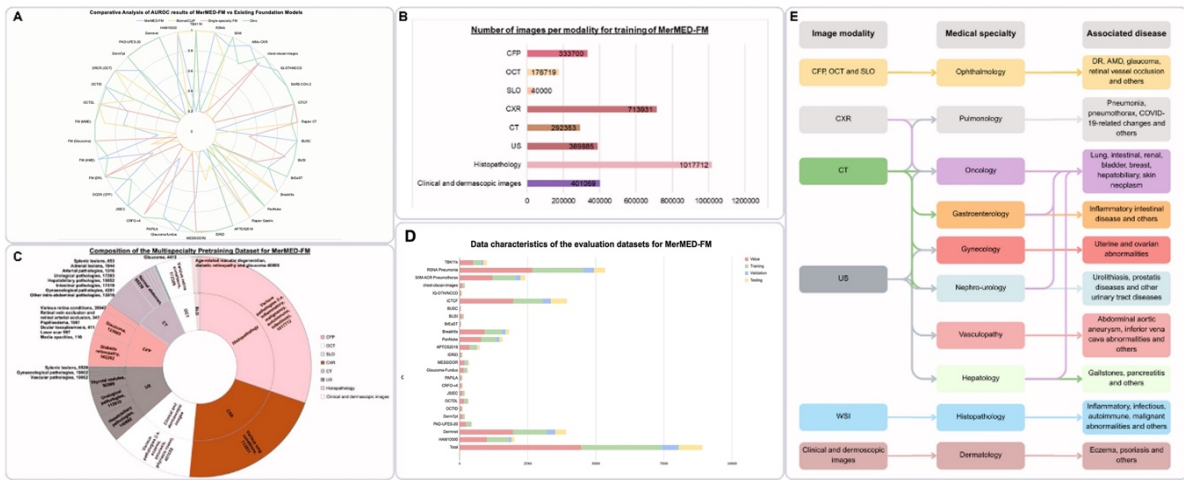

**Figure 1.** Development and Benchmarking of MerMED-FM: Performance, Data Composition, and Clinical Applications. **A.** Radar plot comparing the AUROC of multispecialty FM MerMED-FM against general domain FM (Dino); respective single-specialty FMs (RETFound [OCT and CFP], RadDINO [CXR], USFM [US], UNI [pathology] and PanDerm [dermatology]); multispecialty FMs (BiomedCLIP). Values were normalized using the following formula for clear visual representation:

$$Normalized\ value = 0.2 + \frac{(x - mix(x)) \times 0.8}{\max(x) - \min(x)}$$

**B.** Number of images of pretraining dataset for MerMED-FM per modality represented in a bar graph. **C.** A breakdown of disease types within each modality highlights the diversity of cases used to train the model. **D.** Data characteristics of the fine-tuning datasets for MerMED-FM. **E.** Mapping of imaging modalities to medical specialties and associated diseases for MerMED-FM

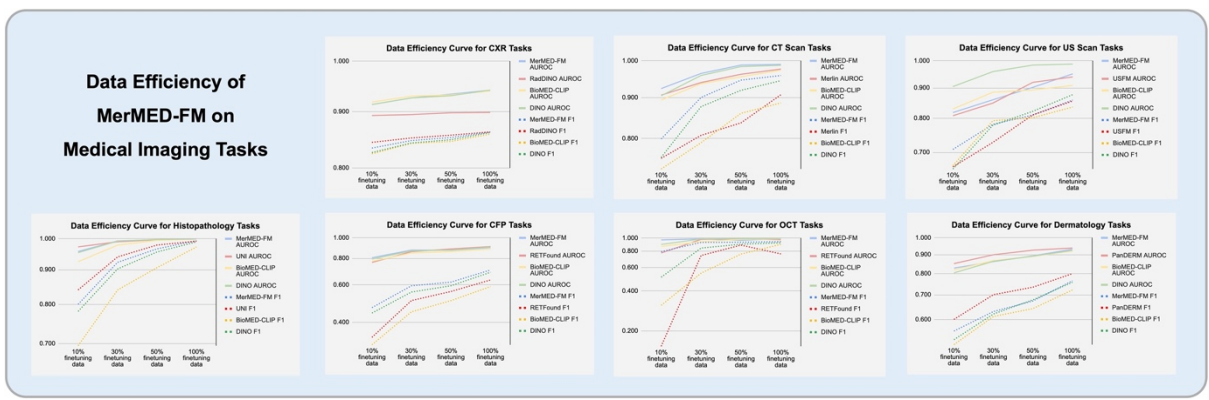

**Figure 2**. Data efficiency of MerMED-FM on medical imaging tasks including (a) CXR = Chest X-Ray; (b) CT = Computed Tomography; (c) US = Ultrasound; (d) histopathology; (e) CFP = Colour Fundus Photography; (f) OCT = Optical Coherence Tomography; (g) dermatology.

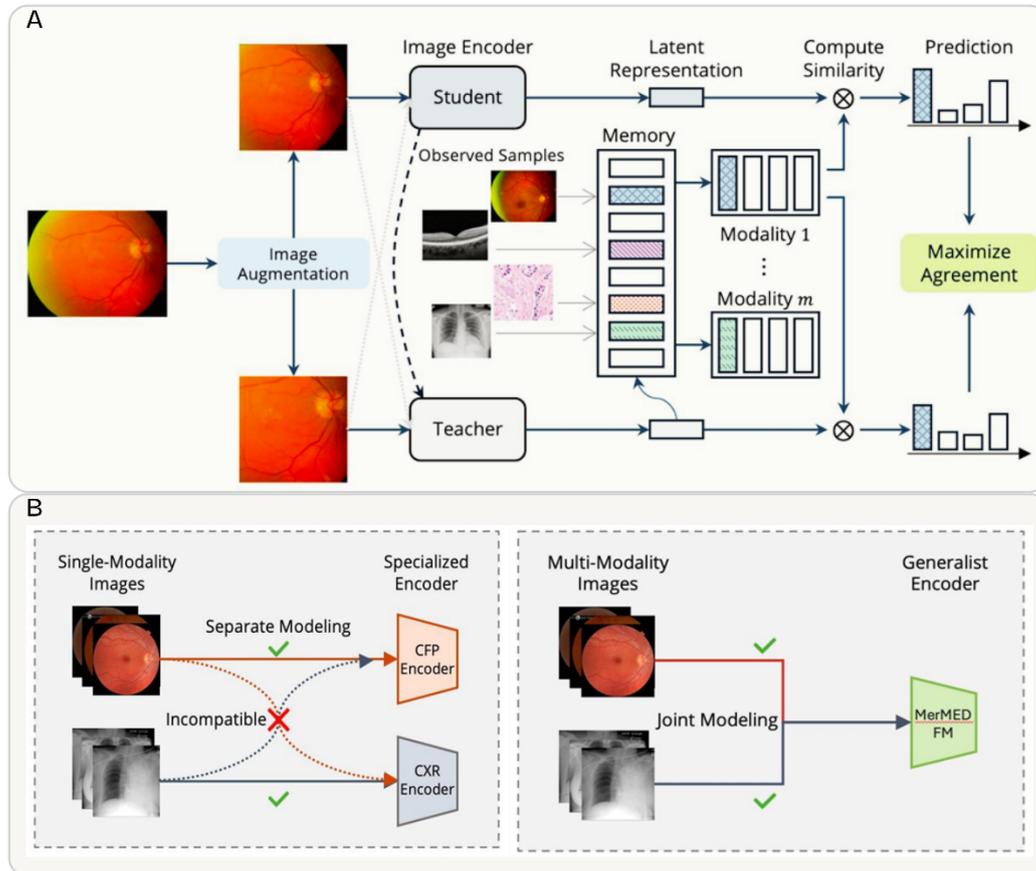

**Figure 3**. Methodological Comparison: MerMED-FM vs. Existing Models. A. Flowchart of MerMED-FM's methodology. A joint-embedding teacher-student architecture using ViT as the backbone. Operating on a FIFO update strategy, the memory module (M) is a non-differentiable container that stores representations generated during training. B. Existing models are predominantly single modality and are modeled separately. MerMED-FM effectively integrates multiple imaging techniques via joint modelling methods. CFP = Color fundus photography; CXR = Chest X-ray

**Tables**

**Table 1.** AUROC, sensitivity, and specificity of MerMED-FM across public and local datasets, organized by disease type, dataset source (public or local), and imaging modality.

Abbreviations: CXR = Chest X-ray; CT = Computed Tomography; US = Ultrasound; CFP = Color Fundus Photography; OCT = Optical Coherence Tomography.

|  | Disease(s) | Dataset | Results of MerMED-FM | | | Results of MerMED-FM by disease type | | | Results of MerMED-FM by public and local dataset | | | Results of MerMED-FM by modality type | | |
|---|---|---|---|---|---|---|---|---|---|---|---|---|---|---|
|  |  |  | AUROC | Sensitivity | Specificity | Mean AUROC | Mean sensitivity | Mean specificity | Mean AUROC | Mean sensitivity | Mean specificity | Mean AUROC | Mean sensitivity | Mean specificity |
| **CXR** | Public datasets |  |  |  |  |  |  |  |  |  |  | 0.908 | 78.9% | 86.9% |
|  | Tuberculosis | TBX11K | 0.999 | 99.2% | 99.7% | 0.999 | 99.2% | 99.7% | 0.941 | 85.7% | 85.9% |  |  |  |
|  | Pneumonia | RSNA | 0.890 | 75.8% | 75.8% | 0.890 | 75.8% | 75.8% |  |  |  |  |  |  |
|  | Pneumothorax | SIIM | 0.934 | 82.1% | 82.1% | 0.934 | 82.1% | 82.1% |  |  |  |  |  |  |
|  | Local dataset |  |  |  |  |  |  |  |  |  |  |  |  |  |
|  | Various respiratory diseases (pneumothorax, pneumonia, pulmonary edema, pulmonary mass, rib fracture) | AIMx-CXR | 0.808 | 58.4% | 89.8% | 0.808 | 58.4% | 89.8% | 0.808 | 58.4% | 89.8% |  |  |  |

| Modality | Category | Dataset | | | | | | | | | | | | | |
|---|---|---|---|---|---|---|---|---|---|---|---|---|---|---|---|
| **CT** | Public datasets | | | | | | | | | | | 0.990 | 95.5% | 98.6% | |
| | Lung carcinoma | chest-ctscan-images | 0.988 | 97.4% | 98.9% | 0.981 | 92.4% | 98.2% | 0.990 | 95.5% | 98.6% | | | | |
| | | IQ-OTHNCCD | 0.974 | 87.4% | 97.6% | | | | | | | | | | |
| | COVID-19 | SARS-COV-2 | 0.999 | 98.7% | 98.7% | 0.999 | 98.6% | 99.0% | | | | | | | |
| | | iCTCF | 0.999 | 98.5% | 99.3% | | | | | | | | | | |
| **US** | Public datasets | | | | | | | | | | | 0.951 | 86.0% | 88.5% | |
| | Breast cancer | BUSC | 1.000 | 98.7% | 98.7% | 0.951 | 86.0% | 88.5% | 0.951 | 86.0% | 88.5% | | | | |
| | | BUSI | 0.973 | 84.7% | 92.2% | | | | | | | | | | |
| | | BrEaST | 0.878 | 74.6% | 74.6% | | | | | | | | | | |
| **Histopathology** | Public datasets | | | | | | | | | | | 1.000 | 99.5% | 99.8% | |
| | Colorectal adenocarcinoma | CRC-VAL-HE-7K | 1.000 | 99.9% | 100.0% | 1.000 | 99.9% | 100.0% | 1.000 | 99.5% | 99.8% | | | | |
| | Breast cancer | BreakHis | 1.000 | 99.5% | 99.5% | 1.000 | 99.5% | 99.5% | | | | | | | |
| | Various tissue types (i.e. breast, colon, adrenal gland, esophagus, etc.) | PanNuke | 1.000 | 99.1% | 100.0% | 1.000 | 99.1% | 100.0% | | | | | | | |
| **CFP** | Public datasets | | | | | | | | | | | | | | |
| | DR | APTOS2019 | 0.918 | 66.3% | 95.8% | 0.852 | 55.6% | 90.8% | 0.899 | 68.9% | 92.4% | 0.894 | 64.3% | 88.8% | |
| | | IDRiD | 0.768 | 39.9% | 86.9% | | | | | | | | | | |

| | | | | | | | | | | | | |
|---|---|---|---|---|---|---|---|---|---|---|---|---|
| | | MESSIDOR2 | 0.870 | 60.4% | 89.5% | | | | | | | |
| | Glaucoma | Glaucoma fundus | 0.957 | 85.3% | 93.8% | 0.897 | 73.9% | 88.1% | | | | |
| | | PAPILA | 0.836 | 62.4% | 82.5% | | | | | | | |
| | Various retinal diseases (i.e. Glaucoma, DR, AMD, CSR) | CRFO-v4 | 0.949 | 81.9% | 98.5% | 0.972 | 83.8% | 99.1% | | | | |
| | | JSIEC | 0.996 | 85.7% | 99.7% | | | | | | | |
| | Local datasets | | | | | | | | | | | |
| | Diabetic macular edema | DCDR (CFP) | 0.820 | 74.3% | 74.3% | 0.820 | 74.3% | 74.3% | 0.886 | 58.0% | 83.8% | |
| | DR | FM (DR) | 0.917 | 55.1% | 92.1% | 0.917 | 55.1% | 92.1% | | | | |
| | AMD | FM (AMD) | 0.937 | 46.5% | 82.1% | 0.937 | 46.5% | 82.1% | | | | |
| | Glaucoma | FM (Glaucoma) | 0.816 | 47.1% | 75.2% | 0.816 | 47.1% | 75.2% | | | | |
| | MMD | FM (MMD) | 0.939 | 66.8% | 95.1% | 0.939 | 66.8% | 95.1% | | | | |
| **OCT** | Public datasets | | | | | | | | | | | |
| | Various retinal diseases (ie. DR, AMD, DME, ERM, | OCTDL | 0.991 | 92.7% | 98.9% | 0.994 | 93.9% | 99.0% | 0.994 | 93.9% | 99.0% | 0.988 | 92.8% | 96.2% |

| Specialty | Category | Dataset | | | | | | | | | | | |
|---|---|---|---|---|---|---|---|---|---|---|---|---|---|
| | | RVO, RAO, macular hole, vitreomacular interface diseases) | | | | | | | | | | | |
| | | OCTID | 0.998 | 95.1% | 99.1% | | | | | | | |
| | Local dataset | | | | | | | | | | | |
| | Diabetic macular edema | DCDR (OCT) | 0.975 | 90.6% | 90.6% | 0.975 | 90.6% | 90.6% | 0.975 | 90.6% | 90.6% | |
| **Dermatology** | Public datasets | | | | | | | | | | | 0.931 | 75.9% | 92.1% |
| | Skin lesions | Derm7pt | 0.845 | 76.7% | 76.7% | 0.913 | 72.5% | 90.0% | 0.931 | 75.9% | 92.1% | |
| | | PAD-UFES-20 | 0.930 | 70.2% | 94.6% | | | | | | | |
| | | Dermnet | 0.962 | 70.6% | 98.8% | | | | | | | |
| | Pigmented skin lesions | HAM10000 | 0.987 | 86.2% | 98.1% | 0.987 | 86.2% | 98.1% | | | | |

**Table 2.** AUROC comparison across published datasets for MerMED-FM, general-domain model DINO, single-modality foundation models, and the multimodal model, BiomedCLIP.

Abbreviations: CXR = Chest X-ray; CT = Computed Tomography; US = Ultrasound; CFP = Color Fundus Photography; OCT = Optical Coherence Tomography; DR = Diabetic Retinopathy; AMD = Age-related Macular Degeneration; CSR = Central Serous Retinopathy; RVO = Retinal Vein Occlusion; RAO = Retinal Artery Occlusion; ERM = Epiretinal Membrane; HCC = Hepatocellular carcinoma.

**Bold and underlined** = best performance; **bold only** = second-best performance.

| | | | MerMED-FM | | | Multispecialty FM (BiomedCLIP) | | | Single-specialty FM | | | General domain FM (Dino) | | |
|---|---|---|---|---|---|---|---|---|---|---|---|---|---|---|
| | Disease(s) | Dataset | AUROC | AUROC confidence interval lower limit | AUROC confidence interval higher limit | AUROC | AUROC confidence interval lower limit | AUROC confidence interval higher limit | AUROC | AUROC confidence interval lower limit | AUROC confidence interval higher limit | AUROC | AUROC confidence interval lower limit | AUROC confidence interval higher limit |
| **CXR** | | | | | | | | | RadDINO | | | | | |
| | | Public dataset | | | | | | | | | | | | |
| | Tuberculosis | TBX11K | **0.999** | 0.999 | 1.000 | **<u>1.000</u>** | 1.000 | 1.000 | **0.999** | 0.999 | 0.999 | **0.999** | 0.999 | 0.999 |
| | Pneumonia | RSNA | **0.890** | 0.889 | 0.891 | **<u>0.891</u>** | 0.891 | 0.892 | 0.793 | 0.785 | 0.802 | 0.888 | 0.885 | 0.890 |
| | Pneumothorax | SIIM | **0.934** | 0.932 | 0.937 | 0.928 | 0.925 | 0.931 | 0.903 | 0.893 | 0.913 | **<u>0.938</u>** | 0.937 | 0.939 |
| | | Local dataset | | | | | | | | | | | | |
| | Overall | AIMx-CXR | **<u>0.865</u>** | 0.808 | 0.802 | 0.760 | 0.752 | 0.768 | **0.807** | 0.796 | 0.818 | 0.788 | 0.783 | 0.793 |
| | Pneumothorax | | **<u>0.880</u>** | 0.865 | 0.894 | 0.788 | 0.759 | 0.818 | 0.830 | 0.784 | 0.876 | **0.850** | 0.831 | 0.869 |

| | | | | | | | | | | | | | |
|---|---|---|---|---|---|---|---|---|---|---|---|---|---|
| | Rib fracture | | **0.773** | 0.750 | 0.795 | 0.722 | 0.706 | 0.737 | 0.737 | 0.711 | 0.762 | **0.742** | 0.717 | 0.768 |
| | Pneumonia | | **0.840** | 0.827 | 0.852 | 0.830 | 0.824 | 0.836 | **0.858** | 0.853 | 0.863 | 0.834 | 0.826 | 0.843 |
| | Pulmonary edema | | 0.835 | 0.818 | 0.851 | **0.853** | 0.839 | 0.867 | **0.858** | 0.848 | 0.869 | 0.850 | 0.839 | 0.862 |
| | Lung mass | | **0.714** | 0.692 | 0.737 | 0.608 | 0.593 | 0.623 | **0.751** | 0.729 | 0.774 | 0.662 | 0.639 | 0.685 |
| **CT** | | | | | | | | | **Merlin** | | | | | |
| | | Public dataset | | | | | | | | | | | | |
| | Lung carcinoma | chest-ctscan-images | **0.988** | 0.985 | 0.990 | 0.975 | 0.970 | 0.979 | 0.981 | 0.977 | 0.985 | **0.990** | 0.986 | 0.994 |
| | | IQ-OTHNCCD | **0.974** | 0.964 | 0.984 | 0.939 | 0.935 | 0.942 | 0.940 | 0.926 | 0.953 | **0.963** | 0.956 | 0.970 |
| | COVID-19 | SARS-COV-2 | **0.999** | 0.999 | 0.999 | 0.981 | 0.979 | 0.983 | 0.987 | 0.986 | 0.989 | **0.997** | 0.995 | 0.999 |
| | | iCTCF | **0.999** | 0.999 | 1.000 | **0.999** | 0.999 | 1.000 | **0.999** | 0.998 | 0.999 | **0.999** | 0.999 | 1.000 |
| | | Local dataset | | | | | | | | | | | | |
| | Overall | RAPIERCT | 0.916 | 0.901 | 0.931 | **0.953** | 0.948 | 0.958 | **0.939** | 0.932 | 0.946 | 0.937 | 0.915 | 0.960 |
| | Hepatocellular carcinoma (HCC) | | **0.962** | 0.957 | 0.967 | 0.925 | 0.918 | 0.933 | 0.931 | 0.924 | 0.938 | **0.948** | 0.945 | 0.952 |
| | Liver cyst | | **0.948** | 0.933 | 0.963 | **0.958** | 0.950 | 0.965 | 0.933 | 0.926 | 0.941 | 0.947 | 0.931 | 0.963 |
| | Liver malignanc | | **0.954** | 0.941 | 0.966 | 0.917 | 0.914 | 0.921 | 0.932 | 0.926 | 0.938 | **0.940** | 0.935 | 0.945 |

| | | | | | | | | | | | | | |
|---|---|---|---|---|---|---|---|---|---|---|---|---|---|
| | | y other than HCC | | | | | | | | | | | |
| | | Hemangioma | | **0.960** | 0.936 | 0.983 | 0.947 | 0.936 | 0.958 | 0.945 | 0.939 | 0.952 | <u>**0.961**</u> | 0.952 | 0.970 |
| | | Abscess | | **0.964** | 0.951 | 0.978 | <u>**0.967**</u> | 0.957 | 0.976 | 0.936 | 0.926 | 0.945 | 0.945 | 0.896 | 0.993 |
| | | Focal nodular hyperplasia | | 0.644 | 0.532 | 0.756 | <u>**0.974**</u> | 0.967 | 0.981 | 0.937 | 0.901 | 0.973 | 0.827 | 0.655 | 0.999 |
| | | Benign liver lesion | | 0.981 | 0.961 | 1.001 | **0.986** | 0.976 | 0.996 | 0.958 | 0.939 | 0.977 | <u>**0.995**</u> | 0.992 | 0.997 |
| **US** | | | | | | | | | | **USFM** | | | | | |
| | | | Public dataset | | | | | | | | | | | | |
| | | Breast cancer | BUSC | <u>**1.000**</u> | 1.000 | 1.000 | 0.945 | 0.927 | 0.962 | **0.997** | 0.992 | 1.002 | 0.995 | 0.989 | 1.002 |
| | | | BUSI | **0.973** | 0.968 | 0.979 | 0.960 | 0.955 | 0.965 | 0.966 | 0.963 | 0.969 | <u>**0.975**</u> | 0.971 | 0.980 |
| | | | BrEaST | **0.878** | 0.857 | 0.899 | 0.821 | 0.778 | 0.864 | 0.854 | 0.832 | 0.877 | <u>**0.898**</u> | 0.865 | 0.931 |
| **Histopathology** | | | | | | | | | | **UNI** | | | | | |
| | | | Public dataset | | | | | | | | | | | | |
| | | Breast cancer | BreakHis | <u>**0.999**</u> | 0.999 | 1.000 | <u>**0.999**</u> | 0.998 | 0.999 | **0.998** | 0.997 | 1.000 | <u>**0.999**</u> | 0.999 | 1.000 |
| | | Various tissue types (i.e. breast, colon, adrenal gland, | PanNuke | **0.999** | 0.999 | 1.000 | 0.999 | 0.998 | 1.000 | <u>**1.000**</u> | 1.000 | 1.000 | **0.999** | 0.999 | 1.000 |

| | | | | | | | | | | | | | |
|---|---|---|---|---|---|---|---|---|---|---|---|---|---|
| | esophagus, etc.) | | | | | | | | | | | | |
| | | Local dataset | | | | | | | | | | | |
| | Overall | RAPIER | 0.970 | 0.969 | 0.971 | 0.966 | 0.966 | 0.967 | 0.961 | 0.958 | 0.963 | 0.965 | 0.965 | 0.966 |
| | Gastric intestinal metaplasia | Gastric | **<u>0.992</u>** | 0.992 | 0.992 | 0.990 | 0.989 | 0.990 | 0.989 | 0.988 | 0.989 | **0.991** | 0.990 | 0.993 |
| | Gastric mucosa | | **<u>0.962</u>** | 0.960 | 0.963 | 0.955 | 0.954 | 0.956 | 0.948 | 0.946 | 0.949 | **0.957** | 0.956 | 0.959 |
| | Gastric H pylori | | **<u>0.957</u>** | 0.954 | 0.959 | **0.955** | 0.954 | 0.956 | 0.946 | 0.940 | 0.951 | 0.948 | 0.945 | 0.950 |
| **CFP** | | | | | | | | | **RETFound (CFP)** | | | | | |
| | | Public dataset | | | | | | | | | | | | |
| | DR | APTOS2019 | 0.918 | 0.906 | 0.930 | 0.937 | 0.934 | 0.939 | **<u>0.945</u>** | 0.943 | 0.947 | **0.935** | 0.926 | 0.944 |
| | | IDRiD | 0.768 | 0.753 | 0.783 | **0.787** | 0.770 | 0.805 | **<u>0.818</u>** | 0.808 | 0.829 | 0.769 | 0.753 | 0.785 |
| | | MESSIDOR2 | 0.870 | 0.862 | 0.878 | 0.856 | 0.849 | 0.863 | **0.878** | 0.873 | 0.882 | **<u>0.881</u>** | 0.880 | 0.882 |
| | Glaucoma | Glaucoma fundus | **<u>0.957</u>** | 0.953 | 0.961 | 0.919 | 0.916 | 0.921 | **0.937** | 0.935 | 0.939 | 0.936 | 0.933 | 0.939 |
| | | PAPILA | **0.836** | 0.820 | 0.852 | 0.762 | 0.744 | 0.780 | **<u>0.847</u>** | 0.834 | 0.860 | 0.830 | 0.809 | 0.852 |
| | Various retina diseases (i.e. Glaucoma, DR, AMD, CSR) | CRFO-v4 | 0.949 | 0.941 | 0.957 | 0.943 | 0.936 | 0.949 | **0.955** | 0.947 | 0.963 | **<u>0.987</u>** | 0.977 | 0.997 |
| | | JSIEC | **<u>0.996</u>** | 0.993 | 0.998 | 0.989 | 0.987 | 0.991 | 0.990 | 0.989 | 0.990 | **0.994** | 0.991 | 0.998 |

|   | | Local dataset | | | | | | | | | | | | |
|---|---|---|---|---|---|---|---|---|---|---|---|---|---|---|
|   | Diabetic macular edema | DCDR (CFP) | **0.820** | 0.810 | 0.830 | 0.772 | 0.764 | 0.780 | 0.777 | 0.770 | 0.785 | **0.794** | 0.792 | 0.797 |
|   | DR | FM (DR) | 0.917 | 0.899 | 0.935 | 0.907 | 0.905 | 0.909 | **0.936** | 0.932 | 0.939 | **0.928** | 0.918 | 0.938 |
|   | AMD | FM (AMD) | **0.937** | 0.932 | 0.942 | 0.877 | 0.855 | 0.898 | 0.849 | 0.835 | 0.863 | **0.935** | 0.929 | 0.941 |
|   | Glaucoma | FM (Glaucoma) | **0.816** | 0.808 | 0.823 | 0.810 | 0.807 | 0.812 | **0.837** | 0.832 | 0.842 | 0.814 | 0.811 | 0.817 |
|   | MMD | FM (MMD) | 0.939 | 0.936 | 0.942 | **0.947** | 0.898 | 0.997 | **0.951** | 0.936 | 0.965 | 0.917 | 0.865 | 0.968 |
| **OCT** | | | | | | | | | | **RETFound (OCT)** | | | | |
|   | | Public dataset | | | | | | | | | | | | |
|   | Various retina diseases (ie. DR, AMD, DME, ERM, RVO, RAO, macular hole, vitreomacular | OCTDL | 0.991 | 0.990 | 0.992 | **0.994** | 0.993 | 0.995 | 0.986 | 0.984 | 0.989 | **0.993** | 0.990 | 0.995 |
|   | | OCTID | **0.998** | 0.997 | 0.999 | 0.982 | 0.978 | 0.986 | 0.960 | 0.954 | 0.966 | **0.996** | 0.993 | 0.998 |

| | | | | | | | | | | | | | |
|---|---|---|---|---|---|---|---|---|---|---|---|---|---|
| | interface diseases) | | | | | | | | | | | | |
| | | Local dataset | | | | | | | | | | | |
| | Diabetic macular edema | DRCR (OCT) | **0.975** | 0.971 | 0.979 | **0.967** | 0.963 | 0.971 | 0.898 | 0.974 | 0.979 | **0.967** | 0.961 | 0.972 |
| **Dermatology** | | | | | | | | | **PanDERM** | | | | |
| | | Public dataset | | | | | | | | | | | |
| | Skin lesions | Derm7pt | **0.845** | 0.833 | 0.857 | 0.811 | 0.803 | 0.818 | **0.855** | 0.830 | 0.881 | 0.824 | 0.796 | 0.852 |
| | | PAD-UFES-20 | 0.930 | 0.918 | 0.942 | 0.932 | 0.929 | 0.935 | **0.947** | 0.943 | 0.951 | **0.933** | 0.927 | 0.939 |
| | | Dermnet | **0.962** | 0.961 | 0.963 | 0.957 | 0.955 | 0.958 | **0.970** | 0.968 | 0.972 | 0.958 | 0.956 | 0.959 |
| | Pigmented skin lesions | HAM10000 | **0.987** | 0.983 | 0.992 | **0.986** | 0.985 | 0.987 | 0.981 | 0.974 | 0.988 | 0.983 | 0.977 | 0.988 |

**Table 3.** Comparative performance of MerMED-FM across published and local datasets against single-modality foundation models, the multimodal model BiomedCLIP, and the general-domain model DINO.

Abbreviations: CXR = Chest X-ray; CT = Computed Tomography; US = Ultrasound; CFP = Color Fundus Photography; OCT = Optical Coherence Tomography; DR = Diabetic Retinopathy; AMD = Age-related Macular Degeneration; CSR = Central Serous Retinopathy; RVO = Retinal Vein Occlusion; RAO = Retinal Artery Occlusion; ERM = Epiretinal Membrane; HCC = Hepatocellular carcioma.

**Bolded** = statistically significant, i.e, p value < 0.05

| | Disease(s) | Dataset | Comparator model | Mean difference between MerMED-FM and the comparator | Percentage difference between MerMED-FM and comparator (%) | T statistic | p value | Cohen's D |
|---|---|---|---|---|---|---|---|---|
| | | | | **Public datasets** | | | | |
| **CXR** | | | | | | | | |
| | Tuberculosis | TBX11K | RadDINO | 2.80E-04 | 0.0280 | 1.14 | 0.318 | 0.637 |
| | | | BiomedCLIP | -3.80E-04 | -0.0380 | -1.35 | 0.249 | -0.928 |
| | | | Dino | 2.20E-04 | 0.0220 | 0.626 | 0.565 | 0.495 |
| | Pneumonia | RSNA | RadDINO | 0.0968 | 12.2 | 29.8 | **7.53E-06** | 20.0 |
| | | | BiomedCLIP | -0.00148 | -0.166 | -2.12 | 0.102 | -1.64 |
| | | | Dino | 0.00244 | 0.275 | 4.16 | **0.0142** | 1.39 |
| | Pneumothorax | SIIM | RadDINO | 0.0312 | 3.45 | 7.92 | **0.00138** | 5.20 |
| | | | BiomedCLIP | 0.0063 | 0.677 | 13.1 | **1.99E-04** | 2.66 |
| | | | Dino | -0.00374 | -0.399 | -4.98 | **0.00759** | -2.45 |
| **CT** | Lung carcinoma | chest-ctscan-images | Merlin | 0.00640 | 0.652 | 3.92 | **0.017** | 2.52 |
| | | | BiomedCLIP | 0.975 | 0.970 | 0.98 | 0.878 | 0.948 |
| | | | Dino | -0.00226 | -0.228 | -1.86 | 0.137 | -0.935 |
| | | IQ-OTHNCCD | Merlin | 0.0339 | 3.61 | 6.02 | **3.83E-03** | 3.51 |
| | | | BiomedCLIP | 0.939 | 0.93 | 0.94 | **7.46E-01** | 0.951 |
| | | | Dino | 0.0105 | 1.09 | 3.10 | **0.0363** | 1.49 |
| | COVID-19 | SARS-COV-2 | Merlin | 0.0116 | 1.17 | 27.4 | **1.05E-05** | 15.3 |
| | | | BiomedCLIP | 0.981 | 0.979 | 0.98 | 0.933 | 0.93 |
| | | | Dino | 0.00236 | 0.237 | 3.07 | **0.0372** | 1.93 |
| | | iCTCF | Merlin | 7.80E-04 | 0.0781 | 39.0 | **2.58E-06** | 6.37 |
| | | | BiomedCLIP | 9.99E-01 | 0.999 | 1.00 | 0.982 | 0.992 |
| | | | Dino | -6.00E-05 | -0.00600 | -1.00 | 0.374 | -0.460 |
| **US** | Breast cancer | BUSC | USFM | 2.90E-03 | 0.291 | 1.58 | 0.189 | 1.00 |
| | | | BiomedCLIP | 9.45E-01 | 0.927 | 0.96 | 0.906 | 0.906 |
| | | | Dino | 4.64E-03 | 0.466 | 1.97 | 0.120 | 1.25 |
| | | BUSI | USFM | 0.00742 | 0.768 | 3.20 | **0.0328** | 2.00 |
| | | | BiomedCLIP | 0.960 | 0.955 | 0.965 | 0.846 | 0.919 |
| | | | Dino | -0.00200 | -0.205 | -2.56 | 0.0627 | -0.463 |
| | | BrEaST | USFM | 0.0239 | 2.80 | 2.10 | 0.103 | 1.36 |
| | | | BiomedCLIP | 0.821 | 0.78 | 0.864 | 0.785 | 0.785 |
| | | | Dino | -0.0197 | -2.20 | -1.72 | 0.161 | -0.878 |
| **Histopathology** | Breast cancer | BreakHis | UNI | 1.20E-03 | 0.120 | 1.89 | 0.132 | 1.12 |
| | | | BiomedCLIP | 9.99E-01 | 0.998 | 1.00 | 0.978 | 0.978 |
| | | | Dino | 2.60E-04 | 0.026 | 1.44 | 0.223 | 0.949 |
| | | PanNuke | UNI | -2.40E-04 | -0.0240 | -1.25 | 0.278 | -0.793 |
| | | Various tissue types (i.e. breast, colon, adrenal gland, esophagus, etc.) | | BiomedCLIP | 9.99E-01 | 0.998 | 1.00 | 0.966 | 0.999 |
| | | | | Dino | 1.40E-04 | 0.0140 | 1.16 | 0.311 | 0.246 |
| **CFP** | DR | APTOS2019 | RETFound | -0.0272 | -2.88 | -5.68 | **0.00475** | -3.92 |
| | | | BiomedCLIP | -0.0187 | -2.00 | -4.41 | **0.0116** | -2.68 |
| | | | Dino | -0.0175 | -1.87 | -4.38 | **0.0119** | -2.02 |
| | | IDRiD | RETFound | -0.0057 | -0.60 | -1.93 | 0.126 | -0.904 |
| | | | BiomedCLIP | -0.0198 | -2.51 | -2.19 | 0.0932 | -1.51 |
| | | | Dino | -0.0381 | -3.86 | -11.3 | **3.52E-04** | -5.39 |
| | | MESSIDOR2 | RETFound | -0.0077 | -0.88 | -2.15 | 0.0978 | -1.47 |
| | | | BiomedCLIP | 0.0140 | 1.64 | 3.99 | **0.0163** | 2.27 |
| | | | Dino | -0.0111 | -1.26 | -4.30 | **0.0127** | -2.38 |
| | Glaucoma | Glaucoma fundus | RETFound | 0.0200 | 2.14 | 12.6 | **2.26E-04** | 7.40 |
| | | | BiomedCLIP | 0.0381 | 4.14 | 17.7 | **6.04E-05** | 13.5 |
| | | | Dino | 0.0207 | 2.21 | 9.44 | **7.02E-04** | 6.99 |
| | | PAPILA | RETFound | -0.0106 | -1.25 | -2.65 | 0.0573 | -0.913 |
| | | | BiomedCLIP | 0.0741 | 9.72 | 7.84 | **0.00143** | 5.39 |
| | | | Dino | 0.00584 | 0.703 | 0.53 | 0.624 | 0.386 |
| | Various retina diseases (i.e. Glaucoma, DR, AMD, CSR) | CRFO-v4 | RETFound | -0.00574 | -0.601 | -1.9 | 0.126 | -0.90 |
| | | | BiomedCLIP | 0.0064 | 0.683 | 4.6 | **0.0102** | 1.15 |
| | | | Dino | -0.0381 | -3.86 | -11.3 | **3.52E-04** | -5.39 |
| | | JSIEC | RETFound | 0.00626 | 0.633 | 7.24 | **0.00194** | 4.36 |
| | | | BiomedCLIP | 0.00732 | 0.740 | 6.44 | **0.00298** | 4.14 |
| | | | Dino | 0.00162 | 0.163 | 1.61 | 0.183 | 0.680 |
| **OCT** | Various retina diseases (ie. DR, AMD, DME, ERM, RVO, RAO, macular hole, vitreomacular interface diseases) | OCTDL | RETFound | 0.00484 | 0.491 | 5.60 | **0.00499** | 3.01 |
| | | | BiomedCLIP | -0.00290 | -0.292 | -9.26 | **7.55E-04** | -3.68 |
| | | | Dino | -0.00156 | -0.157 | -1.61 | 0.182 | -0.92 |
| | | OCTID | RETFound | 0.03762 | 3.92 | 17.9 | **5.77E-05** | 10.61 |
| | | | BiomedCLIP | 0.01554 | 1.58 | 16.7 | **7.51E-05** | 6.60 |
| | | | Dino | 0.00218 | 0.219 | 2.91 | **0.0438** | 1.38 |
| **Dermatology** | Skin lesions | Derm7pt | PanDERM | -0.00974 | -1.14 | -1.63 | 0.178 | -0.607 |
| | | | BiomedCLIP | 0.0347 | 4.28 | 7.99 | **0.00133** | 4.324 |
| | | | Dino | 0.0215 | 2.60 | 2.47 | 0.0692 | 1.234 |
| | | PAD-UFES-20 | PanDERM | -0.0171 | -1.80 | -3.78 | **0.0195** | -2.44 |
| | | | BiomedCLIP | -0.0016 | -0.18 | -0.48 | 0.655 | -0.24 |
| | | | Dino | -0.0027 | -0.29 | -0.45 | 0.673 | -0.36 |
| | | Dermnet | PanDERM | -0.00786 | -0.810 | -14.3 | **1.38E-04** | -6.38 |
| | | | BiomedCLIP | 0.00554 | 0.579 | 10.3 | **4.95E-04** | 4.46 |
| | | | Dino | 0.00462 | 0.483 | 7.1 | **2.12E-03** | 4.58 |
| | Pigmented skin lesions | HAM10000 | PanDERM | 6.58E-03 | 0.671 | 1.93 | 0.126 | 1.37 |
| | | | BiomedCLIP | 1.68E-03 | 0.170 | 1.16 | 0.309 | 0.620 |
| | | | Dino | 4.66E-03 | 0.474 | 3.42 | **0.0268** | 1.16 |
| | | **Local datasets** | | | | | | |
| **CXR** | Various respiratory diseases (pneumothorax, pneumonia, | AIMx-CXR | RadDINO | 0.00130 | 0.161 | 0.271 | 0.800 | 0.186 |
| | | | BiomedCLIP | 0.0481 | 5.95 | 19.9 | **<0.001** | 8.44 |
| | pulmonary edema, pulmonary mass, rib fracture) | | Dino | 0.0206 | 2.55 | 7.32 | **0.00200** | 4.58 |
| **CT** | Various hepatic conditions (HCC, liver cyst, liver malignancies other than HCC, hemangioma, abscess, focal nodular hyperplasia, benign liver lesion) | RAPIER CT | Merlin | 0.9298 | 0.91 | 0.948 | 0.459 | 0.948 |
| | | | BiomedCLIP | 0.9534 | 0.95 | 0.958 | 0.514 | 0.951 |
| | | | Dino | 0.9374 | 0.92 | 0.960 | 0.674 | 0.958 |
| **Histopathology** | Various gastric tissue (intestinal metaplasia, gastric mucosa, H pylori) | RAPIER Gastric | UNI | 0.0094 | 0.98 | 21.1 | **2.98E-05** | 6.95 |
| | | | BiomedCLIP | 0.00362 | 0.375 | 7.22 | **0.00195** | 5.62 |
| | | | Dino | 0.0046 | 0.48 | 10.1 | **5.36E-04** | 6.31 |
| **CFP** | Diabetic macular edema | DCDR (CFP) | RETFound | 0.0424 | 5.46 | 12.4 | **2.42E-04** | 6.04 |
| | | | BiomedCLIP | 0.0478 | 6.20 | 14.7 | **1.25E-04** | 6.61 |
| | | | Dino | 0.0254 | 3.19 | 8.45 | **0.00107** | 4.33 |
| | DR | FM (DR) | RETFound | -0.0185 | -1.97 | -2.70 | 0.0539 | -1.75 |
| | | | BiomedCLIP | 0.00980 | 1.08 | 1.37 | 0.243 | 0.94 |
| | | | Dino | -0.0112 | -1.20 | -1.31 | 0.260 | -0.94 |
| | AMD | FM (AMD) | RETFound | 0.0888 | 10.5 | 14.1 | **1.48E-04** | 10.6 |
| | | | BiomedCLIP | 0.0609 | 6.95 | 9.35 | **7.29E-04** | 4.83 |
| | | | Dino | 0.00242 | 0.259 | 0.750 | 0.495 | 0.523 |
| | Glaucoma | FM (Glaucoma) | RETFound | -0.0217 | -2.59 | -11.14 | **3.70E-04** | -4.33 |
| | | | BiomedCLIP | 0.0059 | 0.73 | 1.73 | 0.158 | 1.33 |
| | | | Dino | 0.0015 | 0.18 | 0.56 | 0.604 | 0.33 |
| | MMD | FM (MMD) | RETFound | -0.0114 | -1.20 | -2.49 | 0.0675 | -1.33 |
| | | | BiomedCLIP | -0.0081 | -0.86 | -0.48 | 0.658 | -0.288 |
| | | | Dino | 0.0225 | 2.46 | 1.28 | 0.271 | 0.770 |
| **OCT** | Diabetic macular edema | DCDR (OCT) | RETFound | -0.00148 | -0.152 | -0.976 | 0.385 | -0.571 |
| | | | BiomedCLIP | 0.00406 | 0.452 | 0.639 | 0.557 | 0.478 |
| | | | Dino | 0.00798 | 0.826 | 3.011 | **0.0395** | 2.09 |